\documentstyle[epsfig]{jaa}
\begin{document}

\title[Variable QPOs in XTE~J1858+034]
{Variable Quasi Periodic Oscillations during an outburst of the
transient X-ray pulsar XTE~J1858+034 }

\author[U. Mukherjee et al.]
{U. Mukherjee$^{1}$, S. Bapna$^{2}$,
H. Raichur$^{1,3}$, B. Paul$^{1}$, S. N. A. Jaaffrey$^{3}$}

\date{}

\maketitle

\begin{center}
$^{1}$Tata Institute of Fundamental Research,\\
Homi Bhabha Road, Colaba, Mumbai 400005\\~\\
$^{2}$Department of Physics, College of Science,\\
M. L. Sukhadia University, Rajasthan 313001, India\\~\\
$^{3}$Joint Astronomy Program, Indian Institute of Science,
Bangalore-560 012\\~\\
\end{center}

\begin{abstract}

We have investigated the Quasi Periodic Oscillation (QPO) properties of the transient
accreting X-ray pulsar XTE~J1858+034 during the second outburst of this source
in April-May 2004. We have used observations made with the Proportional Counter
Array (PCA) of the Rossi X-ray Timing Explorer (RXTE) during May 14-18 2004, 
in the declining phase of the outburst. We detected the presence of low frequency
QPOs in the frequency range of 140--185 mHz in all the RXTE-PCA observations.
We report evolution of the QPO parameters with the time, X-ray flux, and X-ray
photon energy. Though a correlation between the QPO centroid frequency and the
instantaneous X-ray flux is not very clear from the data, we point out that 
the QPO frequency and the one day averaged X-ray flux decreased with time
during these observations. We have obtained a clear energy dependence of the RMS
variation in the QPOs, increasing from about 3\% at 3 keV to 6\% at 25 keV.
The X-ray pulse profile is a single peaked sinusoidal, with pulse fraction
increasing from 20\% at 3 keV to 45\% at 30 keV.
We found that, similar to the previous outburst, the energy spectrum is well
fitted with a model consisting of a cut-off power law along with an iron
emission line.

\end{abstract}

{\bf keywords} stars: pulsar: individual (XTE~J1858+034) --- X-rays:pulsar

\section{Introduction}

Among the accretion powered X-ray pulsars, Quasi Periodic Oscillations (QPO) were
first discovered in the X-ray flux of the transient source EXO 2030+375 (Angelini et al.
1989).  Since then, QPOs have been detected in the X-ray power density spectra of
almost a dozen accretion-powered pulsars (Psaltis 2004). QPO centre frequencies
as seen in the pulsars vary widely, ranging from 1 mHz to 40 Hz. In the models that
deal with QPOs in X-ray pulsars, a correlation is expected between the QPO centre
and the X-ray flux, which in turn is determined by the mass accretion rate. Hence,
intensity dependent variable QPOs can provide an interesting insight into the physics
of accretion. However, QPOs are in general rare and transient events in X-ray pulsars.

To date, six out of the dozen X-ray Binary Pulsars (XBP) showing QPOs are transient
sources (Psaltis 2004). XTE~J1858+034 is one of the transient XBPs in which QPOs
have been detected. This X-ray binary was discovered with the RXTE All Sky Monitor
(ASM) in 1998 February during a transient outburst (Remillard \& Levine 1998).
Observations made with the Proportional Counter Array (PCA) of the RXTE during the
first outburst of the source showed single peaked strong sinusoidal pulsations with
a period of $\sim$221.0 s (Takeshima et al. 1998, Paul \& Rao 1998). QPOs at 110 mHz
with about 7\% RMS fluctuation were discovered from the same observation (Paul \& Rao 1998).
The energy spectrum was measured to be hard, similar to the spectra of XBPs. From the
transient nature of this source, its hard X-ray spectrum, and presence of pulsations, 
it was suggested to be a Be-X-ray binary (Takeshima et al. 1998) which is also supported
from the optical observations of H$_\alpha$ emission line (Reig, Kougentakis \& Papamastorakis
2004).

The source had a second outburst that was first detected with INTEGRAL in March 2004 and
subsequently again in April, 2004 (Molkov et al. 2004). The RXTE-ASM also detected the
outburst and it was followed up with the RXTE-PCA. We detected the QPO
features again during this second outburst from an observation made on May 14, 2004.
Subsequently, a series of target of opportunity observation of XTE~J1858+034 were made
with RXTE during May 16--18 to study the evolution of the QPO properties. We have investigated
the evolution of the QPO parameters with time, X-ray flux and X-ray photon energy. X-ray
pulsation properties and the X-ray spectrum  were also measured from these observations.
In the following sections we present the X-ray observations, analysis and results and a
discussion based on the results obtained.

\section{Observations}

The observations of XTE~J1858+034 reported here were made during May 14-18, 2004 with the
pointed mode instruments of the RXTE satellite. The RXTE-ASM lightcurve of XTE~J1858+034
is shown in Figure 1 for the period of March-May 2004 with a bin-size of one and a half
days. The ASM light curve clearly shows that during the RXTE pointed observations, the
source was in the decaying phase of this outburst. 

The RXTE satellite carries a large area Proportional Counter Array (PCA), consisting
of five Xenon filled proportional counter units (PCUs) sensitive in the energy range of
2--60 keV with a total effective area of 6250 cm$^{2}$ (Jahoda et al.
1996). It also has a set of high energy crystal scintillation experiment (HEXTE; 15--200
keV; 1600 cm$^{2}$ area), and a continuously scanning all-sky monitor
(ASM; 2--10 keV; 90 cm$^{2}$). A full description of the ASM detector
can be found in Bradt et al. (1993). Here we have used the RXTE-ASM long term light curve of
XTE~J1858+034 available from the beginning of the RXTE mission in 1995 to the present and
data from six pointed observations with RXTE-PCA. The details of the observations used in
the present work have been given in a nutshell in Table 1. In total there were data from
eight orbits of the satellite. Two of the six observations, Obs. B \& F had two orbits
of data each (B1, B2 \& F1, F2 in Table 1). The total good time interval (GTI) with the 
PCA from these six observations was 18.7~ks.
Figure 1 shows the RXTE-ASM long term light curve (bin size of one and a half
days for clarity) along with the background subtracted RXTE-PCA count rates superposed on it.
Though the beginning of the outburst is not clear, the gradual fall in the long term intensity
can be clearly seen from the figure.

\section{Data Analysis and Results}

\subsection {Light curve \& Pulse Profile}

From all the eight segments in the six observations, we extracted the light curves,
each with a bin size of 0.125~s and then applied Barycentre correction on them.
All the individual light curves showed the pulses. The combined background subtracted
light curve of all the six observations binned at the pulse period of $\sim$220 s had an
average count rate of $\sim$260 counts s$^{-1}$ and also had a gradual fall in intensity
with time (see figure 2). The pulse folding and the $\chi^{2}$ maximization method gave
a pulse period of 219.823$\pm$0.005 s within 90$\%$ confidence interval.
A slow transient pulsar is expected to
have a large spin-up rate in a high intensity state like this. However, a small duty
cycle during the five days of observation does not allow us to measure the spin-up rate.
The 2-60 keV X-ray light curve, obtained from the PCA detectors is shown in
Figure 2. This figure is generated with a binsize of 220 s, same as the spin period,
which removes the pulsation related intensity variations. Though the overall
intensity changed with time, there was no significant intensity variation at a 
time scale of a few thousand seconds.

We have folded the energy resolved background subtracted light curves with the pulse period
mentioned above and the resultant pulse profiles are shown in Figure 3 for observation A.
The pulse profiles shown for six out of the eight energy bands are single peaked sinusoidal
in energy bands up to 25-35 keV range with
a pulse fraction (defined as Maximum-Minimum/Maximum) gradually increasing from 20\% at
3 keV to 45\% at 30 keV. From Figure 3, we see that pulsations exist upto 45 keV also 
with a pulse fraction of 35\% in the energy band of 35-45 keV. Pulsations above 45 keV 
is not detectable with the PCA due to low signal to noise ratio and low detection efficiency.

\subsection {Power density spectrum \& related results}

We have generated the Power Density Spectra (PDS) from the 2-60 keV 0.125 s time resolved
light curves from each orbit of the satellite. The light curves were broken into segments 
of length 128 s and the PDS obtained from each of these segments
were averaged to produce the final PDS for data from each orbit of the satellite. A broad QPO 
feature between 100 mHz and 200 mHz was very prominent in all the PDS. The continuum of the
PDS in the frequency range of 7 mHz to 2 Hz fits well with a model consisting of a power
law. The prominent QPO feature is fitted well with a Gaussian model component. In most of
the PDS, there is also an evidence for a second QPO peak, probably a harmonic of the fundamental.
However, we have refrained from quantifying the second peak as it has very low statistics.
In Figure 4 we have shown the power density spectrum from observation C which has a power-law
index of -0.93 and the Gaussian, QPO feature centered at 160 mHz. Evolution of the QPO
frequency and the associated RMS for the eight data segments are shown in Figure 5. While
the QPO frequency decreased with time from 185 mHz on May 14th to about 140 mHz on May 18th,
the RMS variability associated with the QPO did not have any significant change. The RMS
variability in the QPO feature is of the order of 4.0~$\%$ to 3.0~$\%$. In order to
investigate the energy dependence of the QPO feature, we generated PDS in the five energy
bands of 2-5 keV, 5-8 keV, 8-12 keV, 12-18 keV and 18-30 keV respectively for Obs A. We 
did not find any evidence of a QPO beyond the 18-30 keV band though.
The PDS in the different energy bands were more or less identical in shape, and were
fitted with the same model as described above. The RMS variation in the QPO feature in
different X-ray energy bands obtained from observation A is plotted in Figure 6. We have
obtained a clear energy dependence of the RMS variation in the QPOs, increasing from
about 3\% at 3 keV to 6\% at 25 keV.
The QPO centroid frequency is plotted against the 3--30 keV RXTE-PCA flux in Figure 7,
which does not show a clear correlation between the two.

\subsection {Energy spectrum}

For spectral analysis, we used the Standard-2 dataset from PCA which is stored with 16 s
time resolution. We first extracted the source plus background spectra from all the eight
segments of observation. The background spectra were modeled using the {\bf ``pcabackest''}
package provided by the XTE guest observer facility (GOF). On the average three  
PCUs were ON, and data from all the PCUs were added together to produce
the spectrum. Channels corresponding to energy less than 3 keV and greater than 30 keV 
were ignored because of low signal to noise ratio. Although the background simulation
model that we have used takes care of the internal background, emission from an extended
source like the Galactic ridge or the cosmic X-ray background are not included in it.
From a detailed observation of the Galactic ridge obtained using the PCAs
(Valinia \& Marshall 1998), we have estimated that about 1.0~$\%$ of the observed flux can be
accounted for by the Galactic ridge emission and a very small fraction by the cosmic X-ray
background. We have, therefore, explicitly incorporated the Galactic ridge emission 
as a sum of Raymond-Smith plasma and a power-law with appropriate normalizations. 
For fitting the pulse-phase averaged X-ray spectrum of the X-ray we have used a high
energy cut-off power law along with a Gaussian emission line. The pulsar spectrum in
the 3--30 keV range is found to be very hard with a photon index $\sim$1.5, cut-off energy
of $\sim$17 keV and e-folding energy of $\sim$16 keV along with a neutral
absorption with a equivalent Hydrogen column density of
$\sim$10$^{23}$ atoms cm$^{-2}$. The spectrum did not show any 
significant variation over the five days of observations. During these observations,
the X-ray flux in the 3-30 keV range varied between
2.9$\times$ 10$^{-9}$ to 5.5$\times$ 10$^{-9}$ erg cm$^{-2}$ s$^{-1}$ 
The X-ray spectrum from one of the observations is shown in figure 7 along with the
best fit model and the residuals.

\section{Discussion}

QPOs in accretion powered X-ray pulsars are in general explained with either the beat
frequency model or the Keplerian
frequency model. In the former (Alpar \& Shaham 1985),
blobs of matter orbit the Neutron Star (NS) at approximately the 
Keplerian orbital frequency of the inner edge of the accretion disk.
The magnetic field rotates at the frequency of the NS and material inflow
to the NS is modulated at the Keplerian frequency . If $\nu_k$
is the Keplerian frequency of the inner edge of the disk and 
$\nu_s$ be the spin frequency of the NS, then according to the beat frequency model,
the frequency of the QPO ($\nu_{QPO}$) is given by the relation
$\nu_{QPO}$ = $\nu_k$ - $\nu_s$. 
According to the latter model (Van der Klis et al. 1987),
the inner edge of the accretion disk contains structures that modulate the
observed flux by obscuration and gives rise to a QPO at $\nu_k$.
A third model by Shirakawa \& Lai (2002) is known as the magnetic disk precession
model. It attributes the presence of QPOs in 
XBPs to the magnetically driven disk warping/precession, concentrated near 
the inner edge of the disk, at the magnetospheric boundary. The warping and 
precession of the disk arises from the magnetic torques which are present due 
to the interactions between the stellar field and the induced electric
currents in the disk.

Out of the six transient XBPs showing QPOs, luminosity dependent QPO properties
have been studied extensively for the two sources EXO 2030+375 (Angelini et al.
1989) and 3A 0535+262 (Finger, Wilson \& Harmon 1996) and both the beat and Keplerian
frequency models are applicable to a good extent. In several XBPs with QPOs
like Her X-1, Cen X-3, 4U 1626-67, LMC X-4 and SMC X-1, and two transients V 0332+52
and 4U 0115+63, the QPO frequency is lower
than the pulsation frequency (Psaltis 2004). In these sources, the Keplerian
frequency model is not applicable, since a Keplerian rotation frequency of the
inner accretion disk that is smaller than the spin frequency of the neutron star
will result into centrifugal inhibition of accretion. Again, for V 0332+52, the beat
frequency model may also not be viable since the value of the  magnetospheric boundary 
calculated from the QPO properties and from the observed luminosity do not
agree (Takeshima et al. 1994). In the Low Mass XBP GRO J1744-28, both these models
are eliminated since in this source also $\nu_{QPO} <  \nu_s$ and a flux change by a
factor of $\sim$7.5 did not have any corresponding change in QPO frequency.

The magnetic disk precession
model (Shirakawa \& Lai 2002) can explain the very low frequency QPOs
in the XBPs, like in 4U 1626-67 (1 mHz) and may also explain the lowest frequency QPOs
detected in some of the other XBPs (Her X-1, LMC X-4, and 4U 0115+63).
While this model robustly accounts for QPOs with frequencies of the
order of 1 mHz, the QPO frequency of XTE~J1858+034 is two orders of magnitude higher
and in this model, the production of QPOs at higher frequency depends on the details of the
physics at the inner edge of the disk. Applicabilty of the magnetic disk precession
model is therefore uncertain in this source.

It has already been mentioned that both the beat frequency and the Keplerian frequency
models of QPOs predict a systematic increase of QPO frequency with the mass-accretion rate,
which can be investigated by looking at the change of QPO frequency with the X-ray flux.
This assumption is generally applicable to transient XBPs, in which
the relation between X-ray luminosity and spin-up torque is well understood. The
spin-up torque is in turn related to the mass accretion rate, inner radius of the
accretion disk and rotation frequency of the inner disk. As the transient XBPs
experience a wide range of mass-accretion rates, these are ideal objects to understand
the QPO phenomena. In the case of XTE~J1858+034
during the present outburst, we have detected changes in the QPO frequency over a
span of five days. There is no discernible correlation between the QPO frequency and
the instantaneous X-ray flux measured from the PCA data, at least during the observation
duration (Figure 7). However, as seen from the RXTE-ASM lightcurve, there is a gradual decrease
in the overall one and a half day averaged X-ray intensity and a gradual decrease in the 
QPO frequency (from 185 mHz to 140 mHz, Figure 5) is also present during these observations.
It may be possible that there is a correlation between the QPO frequency with the overall
X-ray intensity or mass accretion rate which is not reflected in the short term.
We note that in the RXTE-PCA observations of XTE~J1858+034 made during the 1998 outburst,
the source had a count rate (normalised for per detector unit) which was about a factor
of 2.5--3.0 times less compared to the observations reported here. A smaller QPO frequency
of 110 mHz during the 1998 ourburst (Paul \& Rao 1998) compared to the 140-185 mHz
frequency during the 2004 outburst supports an inner accretion disk origin of the QPOs.

Absorption of X-rays by structures in the accretion disk should not be effective in
producing hard X-ray QPOs and this process should definitely be more effective
at lower X-ray energy.  The detection of QPOs upto an energy of 30 keV and energy
dependence of the RMS variability of the QPOs (Figure 6) shows that the Keplerian
frequency model is unlikely to be applicable in the case of XTE~J1858+034.
Considering the strong energy dependence of the QPO rms detected both during the
1998 and the 2004 outbursts, and the possibility of a correlation between the QPO
frequency with the overall X-ray intensity as discussed above, the beat frequency
model seems to be more suitable in the case of XTE~J1858+034.

Spectral parameters of XTE~J1858+034 obtained from the RXTE-PCA observations during the
second outburst in 2004 are similar to those measured during the previous outburst (Paul
\& Rao 1998). The iron emission line has an equivalent width in the range of 125 eV
to 200 eV and the line centre energy is in the range of 6.3--6.7 keV. With the moderate
resolution spectral data from the RXTE-PCA detectors we cannot conclude about the emission
line characteristics, but it is likely that the emission line is from neutral or
lowly ionized iron in the circumstellar material. The column density measured during
the present observations is in the range of
(7.7 -- 12.3)$\times$ 10$^{22}$ atoms cm$^{-2}$, which is higher
than the same during the previous outburst. As the absorption column density depends
on the mass loss history of the companion star prior to the outburst, a change in the
column density is quite likely. We note that the iron line equivalent width and the
absorption column density is consistent with a scenario in which the iron line is
produced in a spherical shell of material which also causes the low energy X-ray
absorption.

\section*{Acknowledgements}
This research has made use of data obtained from the High Energy Astrophysics
Science Archive Research Center (HEASARC), provided by NASA's Goddard Space
Flight Center. UM and HR would like to
acknowledge the Kanwal Rekhi Scholarship of TIFR Endowment Fund for partial
financial support. We also thank Parag Shah for his help in system administration.

\section*{References}
Alpar, M.,A., and J., Shaham, 1985, {\it Nature}, {\bf 316}, 239 \\
Angelini, L., Stella, L., \& Parmar, A.,N., 1989, {\it ApJ}, {\bf 346}, 906 \\
Bradt, H., V., Rothschild, R., E., $\&$ Swank, J., H., 1993, {\it A$\&$AS}, {\bf 97}, 355\\
Finger, M., H., Wilson, R., B., \& Harmon, B., A., 1996, {\it ApJ}, {\bf 459}, 288\\
Jahoda, K., Swank, J., H., Giles, A., B., et.al., 1996, in: Siegmund O.H.W., Gummin M.A.
(eds.) EUV, X-Ray and Gamma-Ray Instrumentation for Astronomy VII. {\bf SPIE 2808, p. 59} \\
Molkov, S.,V. et al., 2004, {\it ATel}, {\bf 274}, 1M \\
Paul, B., and Rao, A.,R., 1998, {\it A\&A}, {\bf 337}, 815 \\
Psaltis, D., {\bf astro-ph/0410536} \\
Remillard, R., and Levine, A., 1998, {\it IAUC 6826} \\
Reig, P., Kougentakis, T., \& Papamastorakis, G., 2004, {\it ATel}, {\bf 308} \\
Shirakawa, A., \& Lai, D., 2002, {\it ApJ}, {\bf 565}, 1134 \\
Takeshima, T., Dotani, T., Mitsuda, K., \& Nagase, F., 1994, {\it ApJ}, {\bf 436}, 871\\
Takeshima, T., Corbet, R., H., D., Marshall, F., E., Swank, J., H.,
\& Chakrabarty, D., 1998, {\it IAUC 6826} \\
Valinia, A., and Marshall, F.,E., 1998, {\it ApJ}, {\bf 505}, 134\\
Van der Klis et al., 1987, {\it ApJ}, {\bf 313}, {\bf L19} \\
\clearpage

\begin{table}
\caption{Observation Log}
\vskip 0.5cm
\begin{tabular}{cccccc}
\hline
\\
Observations & Date &Start Time & Stop Time & Exposure & Count Rate \\
& & (UT) & (UT) & (s) & (s$^{-1}$)
\\
\hline
\\
A &  2004-05-14 & 11:30 &  12:26 & 3344 & 365 \\
\\
B1 & 2004-05-16 & 11:14 & 11:42 & 1665 & 422 \\
\\
B2 & 2004-05-16 & 13:54 & 14:48 & 3213 & 428 \\
\\
C &  2004-05-16 & 18:47 & 19:21 & 2017 & 399 \\
\\
D &  2004-05-16 & 20:24 & 20:45 & 1225 & 408 \\
\\
E &  2004-05-16 & 23:41 & 00:05 & 1403 & 331 \\
\\
F1 & 2004-05-18 & 13:11 & 14:06 & 3297 & 279   \\
\\
F2 & 2004-05-18 & 14:43 & 15:25 & 2517 & 277   \\

\hline
\hline\\
\end{tabular}
\end{table}

\begin{figure}
\vskip 10. cm
\includegraphics{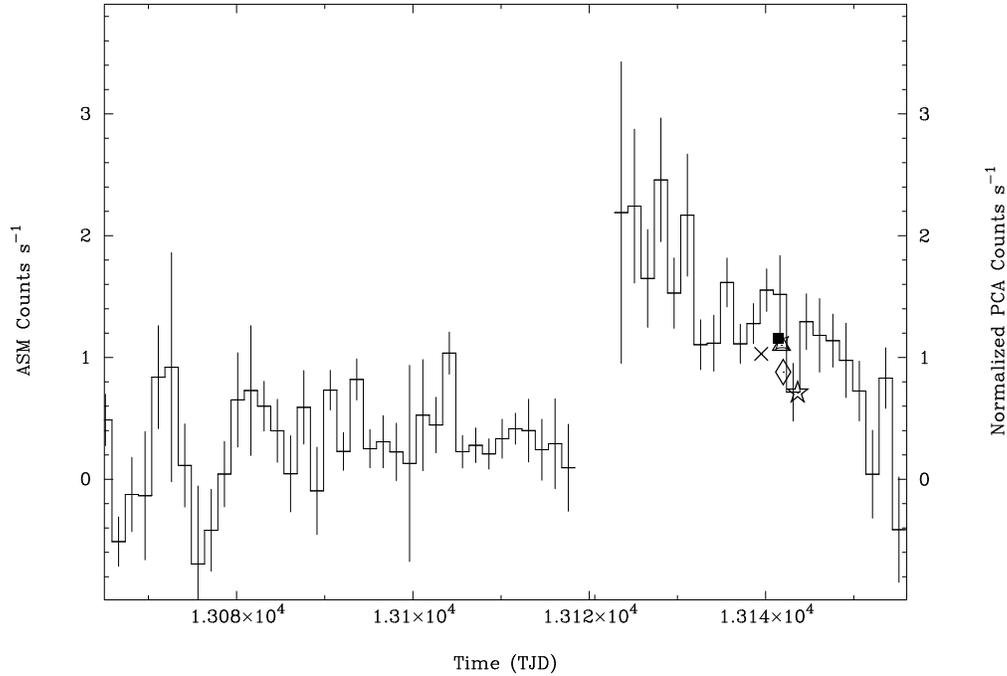}
\caption{The section of the RXTE-ASM long term light curve
depicted between March 1 to May 31, 2005, showing the 2004 outburst in detail. 
The bin size is one and a half days.
The six PCA observations are shown (with different markers) with their respective  
background subtracted count rates (normalized by the mean count rate).}
\end{figure}

\begin{figure}
\vskip 10. cm
\includegraphics{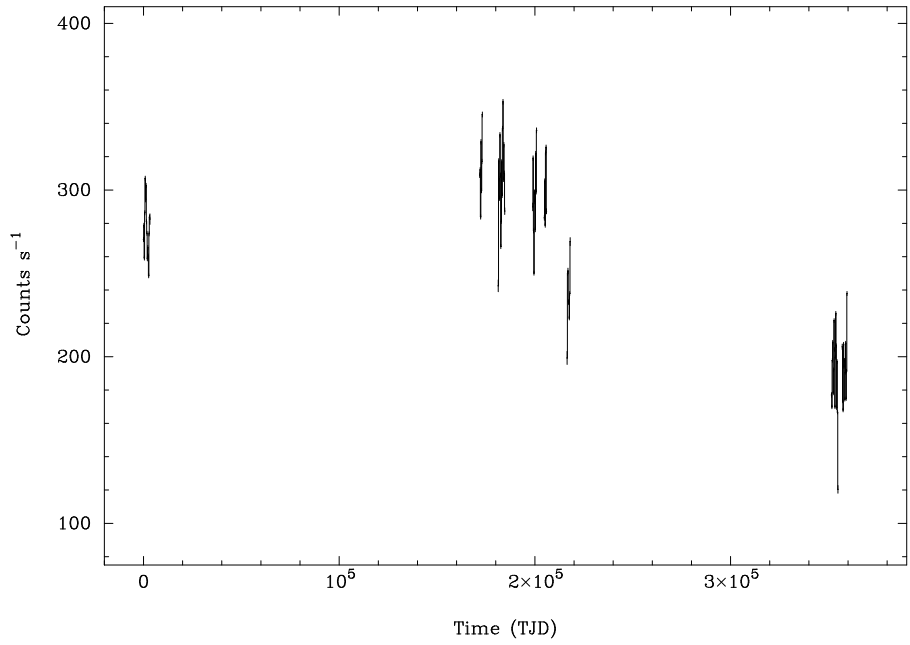}
\caption{The RXTE-PCA background subtracted light curve in the energy band of 2-60 keV
obtained from the 2004 observations is shown here with a binsize of 220 s.}
\end{figure}

\clearpage
\begin{figure}
\vskip 18. cm
\includegraphics{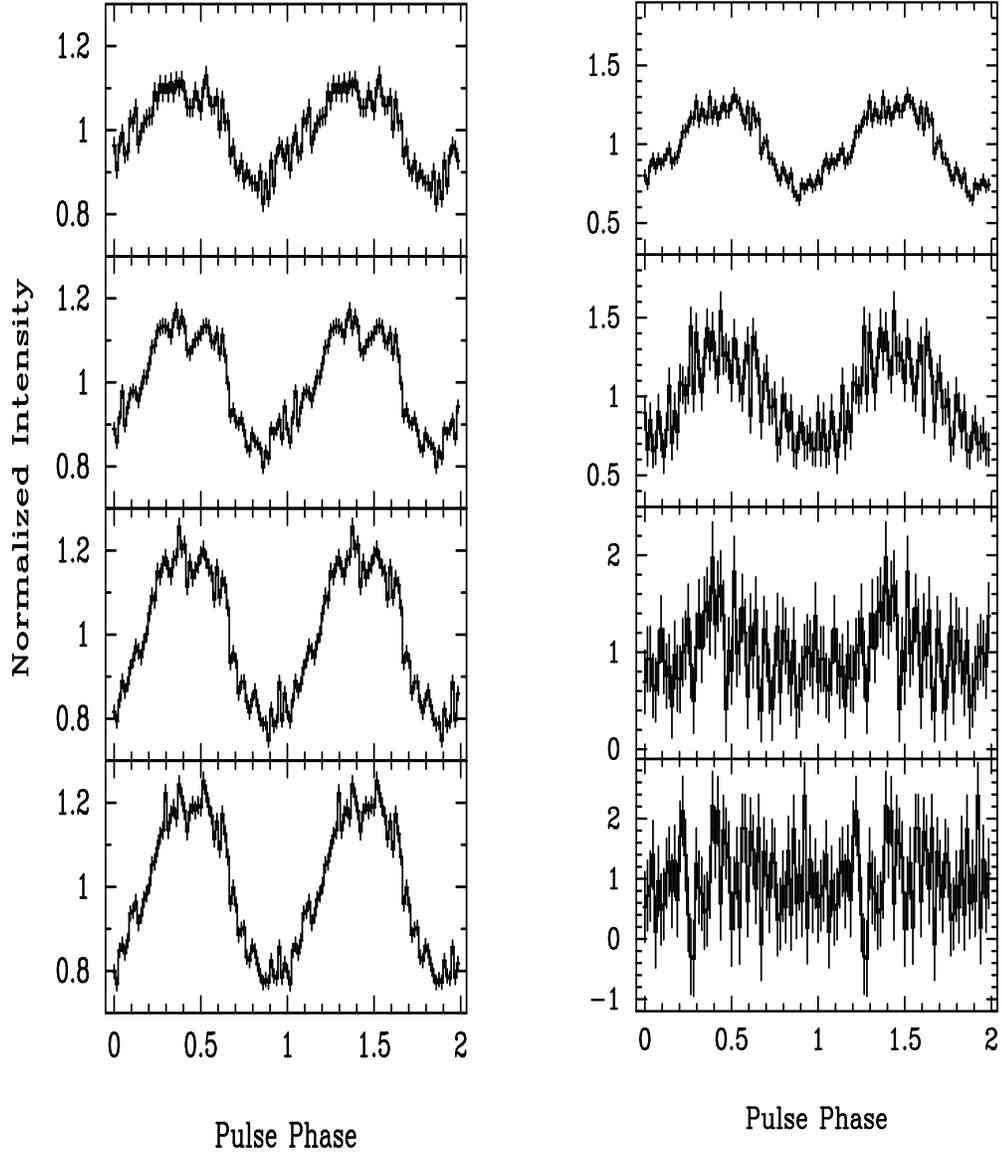}
\caption{The nearly sinusoidal pulse profiles of XTE~J1858+034 folded at a period 
of 219.8 s are shown (left, from the top) for  2-5 keV (first panel), 5-8 keV (second panel),
8-12 keV (third panel) and 12-18 keV (fourth panel). In the right the same is shown 
for 18-25 keV (top panel), 25-35 keV (second panel), 35-45 keV (third panel) and 45-60 keV
(bottom panel) energy bands. The bottom-right panel shows that pulsations are not detectable
with the PCA above 45 keV.}
\end{figure}

\clearpage
\begin{figure}
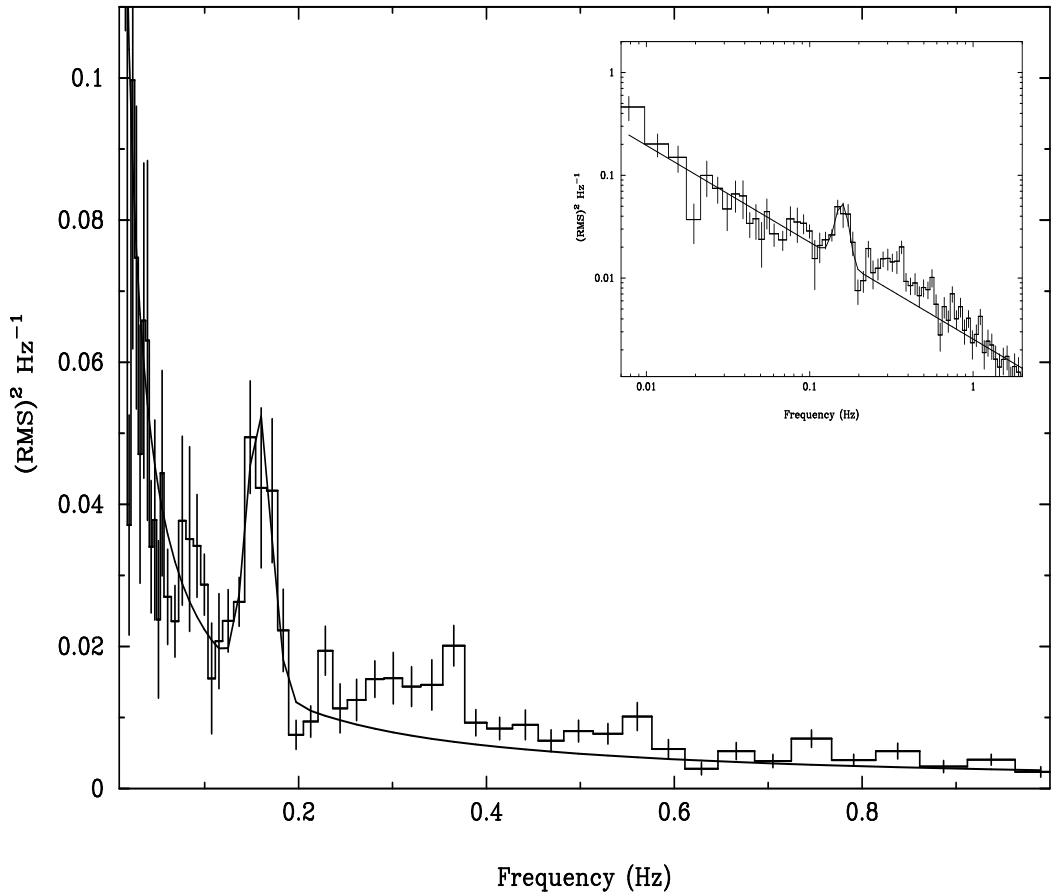

\vskip 14. cm
\includegraphics{fig4a.ps}
\includegraphics{fig4b.ps}
\caption{The power density spectrum of XTE~J1858+034 generated from the 0.125 s
binned lightcurve of Obs. C over the entire energy band of the PCA (in linear scale). 
The line represents the best fitted model in the frequency range of 0.007--2.0 Hz (shown 
for a restricted range to highlight the QPO feature)
consisting of a power law and a Gaussian centred at the QPO frequency.
The figure in the inset is a log-log plot of the same fit.}
\end{figure}

\clearpage
\begin{figure}
\vskip 18. cm
\includegraphics{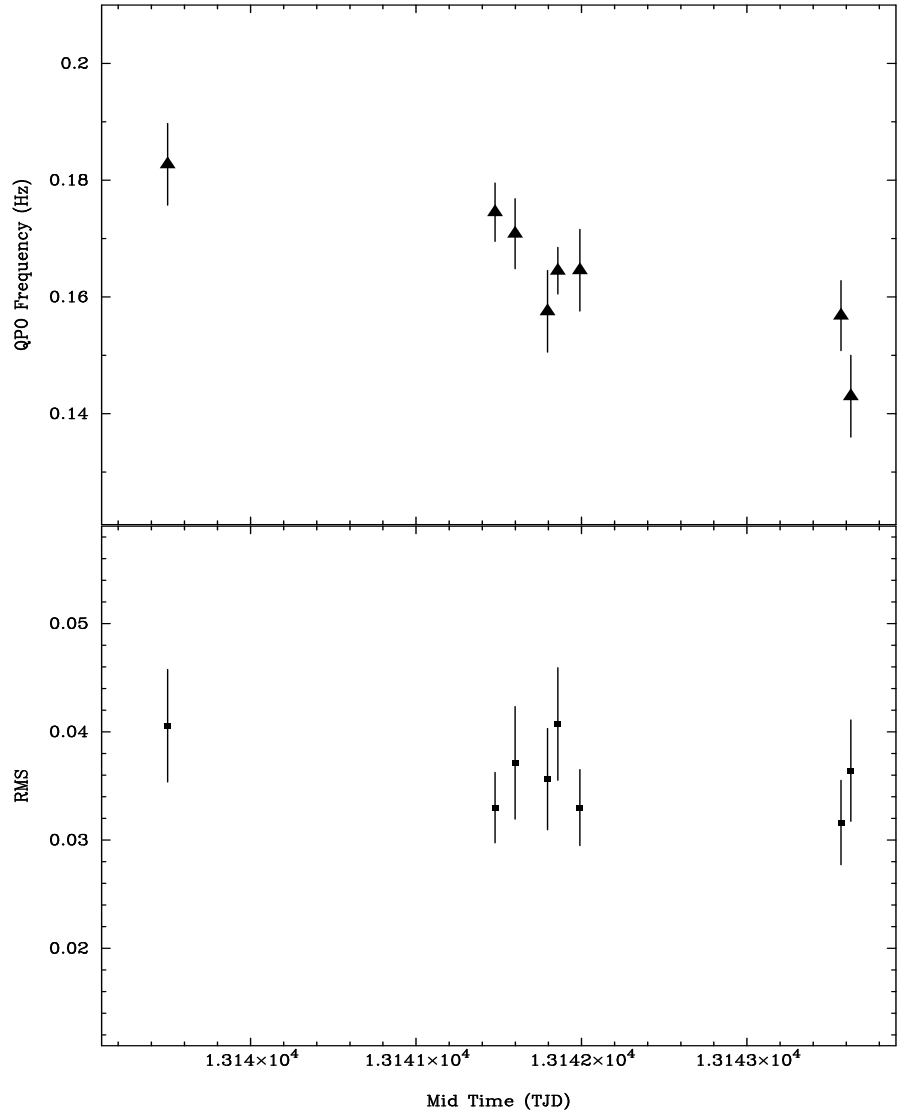}
\caption{The QPO frequency and the RMS variation in the QPO feature
are shown as function of time in the top and the bottom panels
respectively.}
\end{figure}

\clearpage
\begin{figure}
\vskip 10. cm
\includegraphics{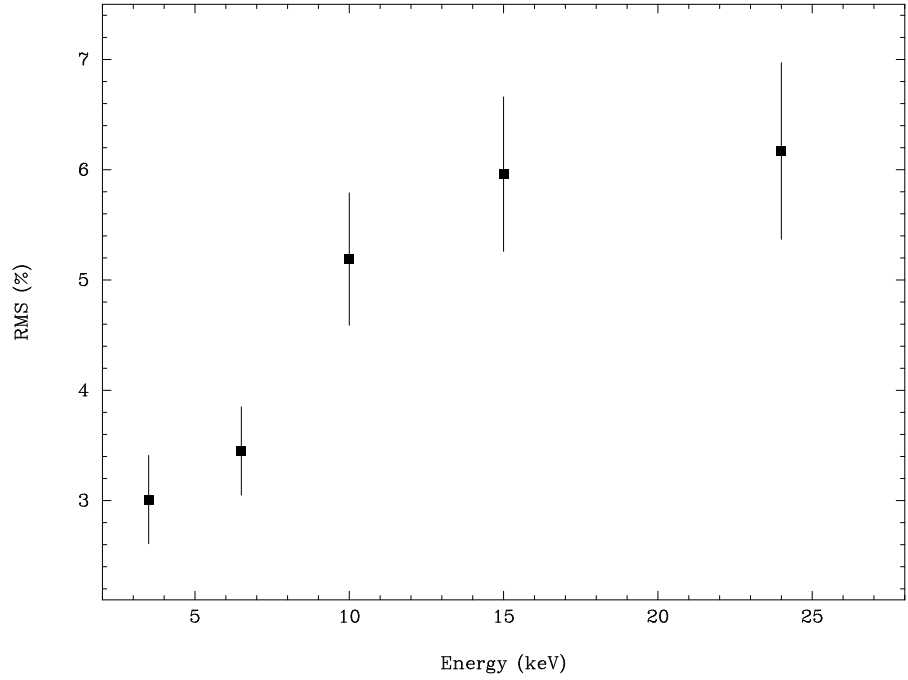}
\caption{The RMS variation (in percentage) in the QPO feature determined from
observation A is shown here as a function of the energy of the X-ray photons.} 
\end{figure}

\clearpage
\begin{figure}
\vskip 10. cm
\includegraphics{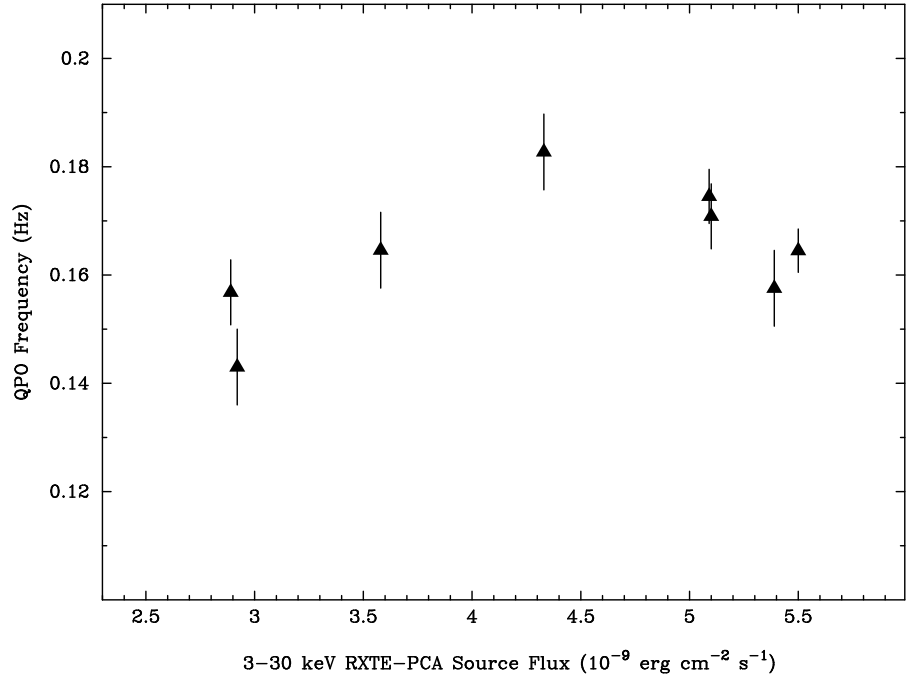}
\caption{The QPO frequency measured from each segment of the RXTE-PCA light
curve are shown here against the 3-30 keV X-ray flux.}
\end{figure}

\clearpage
\begin{figure}
\vskip 10. cm
\includegraphics{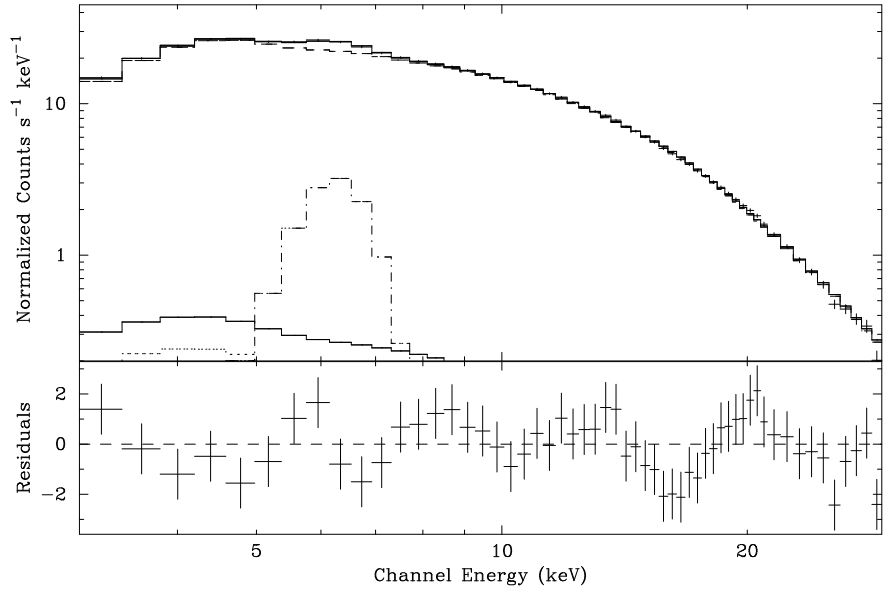}
\caption{The X-ray spectrum obtained from observation B is shown here along with the
best fit model components including a cut-off power law, a Gaussian emission
line, and the additive background components as histograms. The residuals to the
best fit model are shown in the lower panel in unit of their contribution to the
total $\chi^{2}$.}
\end{figure}

\end{document}